\documentclass{article}
\usepackage{spconf,amsmath,graphicx}
\usepackage{booktabs}
\usepackage{siunitx}
\usepackage{amsmath}
\usepackage{amssymb}
\usepackage{mathtools}
\usepackage{amsthm}
\usepackage{multirow}

\title{MLP-ASR: Sequence-length agnostic all-MLP architectures\\ for speech recognition}
\name{Jin Sakuma$^{1,2}$\thanks{Work done when the first author was an intern at LINE corporation.}, Tatsuya Komatsu$^{1}$, Robin Scheibler$^1$}
\address{$^{1}$LINE Corporation, Japan \\ $^{2}$Waseda University, Japan}
\begin{document}
%
\maketitle
\begin{abstract}
We propose multi-layer perceptron (MLP)-based architectures suitable for variable length input.
MLP-based architectures, recently proposed for image classification, can only be used for inputs of a fixed, pre-defined size.
However, many types of data are naturally variable in length, for example, acoustic signals. 
We propose three approaches to extend MLP-based architectures for use with sequences of arbitrary length. 
The first one uses a circular convolution applied in the Fourier domain, the second applies a depthwise convolution, and the final relies on a shift operation. 
We evaluate the proposed architectures on an automatic speech recognition task with the Librispeech and Tedlium2 corpora.
The best proposed MLP-based architectures improves WER by 1.0 / 0.9\%, 0.9 / 0.5\% on Librispeech dev-clean/dev-other, test-clean/test-other set, and 0.8 / 1.1\% on Tedlium2 dev/test set using 86.4\% the size of self-attention-based architecture.
\end{abstract}
\begin{keywords}
speech recognition, connectionist temporal classification, non-autoregressive, multi-layer perceptron
\end{keywords}
\section{Introduction}
Self-attention, the well-known building block of Transformer~\cite{vaswani2017attention}, appeared in natural language processing (NLP) where it caused a breakthrough. 
Soon enough, it propagated to the fields of computer vision and automatic speech recognition (ASR).
In particular, recent end-to-end ASR systems based on self-attention architecture, e.g., Transformer~\cite{karita2019comparative} and Conformer~\cite{gulati2020conformer}, provide state-of-the-art performance.

\begin{figure}
    \begin{center}
    \includegraphics{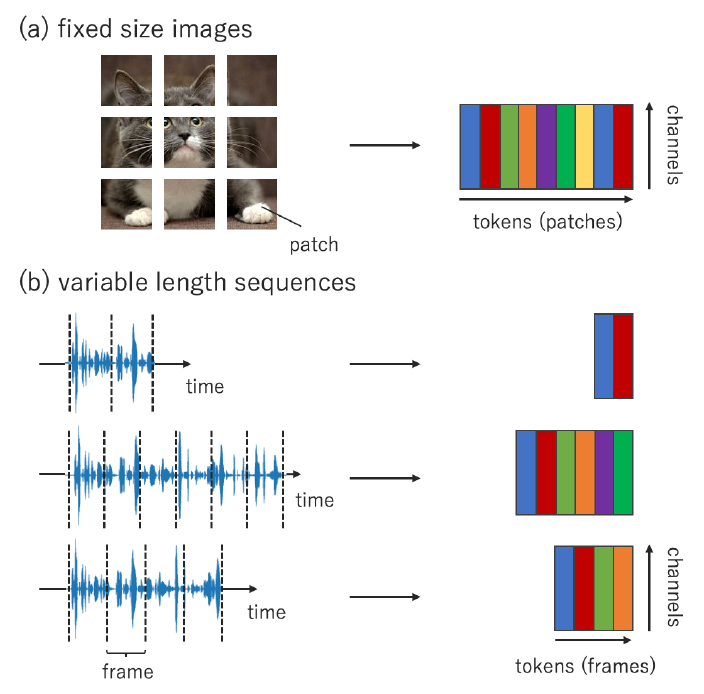}
    \end{center}
    \caption{Illustration of the difference between patching for fixed-size images (a) and variable length sequences (b). We rename patches as \textit{tokens} to adapt to the semantics of sequences.}
    \label{fig:patching}
\end{figure}

Recently, several architectures entirely based on MLP have been proposed in the area of computer vision.
MLP-based architectures have a simple structure, and achieve performance competitive with that of the Transformer-based architectures, despite having fewer parameters and lower computational complexity.
They split an image into patches and reshape it into a $(\text{channels} \times \text{patches})$ matrix used as input.
An illustration of the process, and its analog for variable length sequences, is shown in Figure~\ref{fig:patching}.
MLP-based architecture such as MLP-Mixer ~\cite{Tolstikhin21mlpmixer} and gMLP ~\cite{Liu21gmlp} consist of MLP across the channel dimension and MLP across the patch dimension (also referred to as spatial dimension in computer vision tasks).
MLP across the channel dimension mixes information between channels like a typical feed-forward network (FFN). 
MLP across the patches dimension mixes information between patches. 
All these different works demonstrate that this process capture sufficient information and that self-attention is not always necessary.


For acoustic data, the input sequence is typically first split into contiguous or overlapping blocks and transformed to some kind of frequency domain representation, e.g. a Mel-spectrogram.
Then, the time and frequency dimensions are analogous to patch and channel, respectively, for images.
We adopt the terminology of "token" to describe the vector of channel values at a given time.
It is more apt for sequences and consistent with that used in the MLP-Mixer and gMLP works.
Now, in contrast to images, sequences will produce inputs with a variable number of tokens, making it impractical to apply an MLP directly to the token dimension.
Indeed, due to the fully connected layers, MLPs are not shift invariant, i.e., shifted inputs will produce different outputs.
This is undesirable for sequences. For example, we would like an ASR system to output the same transcript, regardless of when the speech starts in the sequence.

In this paper, we propose three approaches that allow variable length inputs and are shift invariant. 
First, the input sequences are broken down into contiguous chunks that we call tokens, as explained earlier and shown in Figure~\ref{fig:patching}.
Building on previous approaches, we propose three new token mixing units.
In the first one, Fourier MLP (F-MLP), we use the Fourier transform and apply filters in the frequency domain.
This way, sequences of any length can be treated uniformly.
This operation is equivalent to circular depthwise convolution.
It is locally MLP-like along the token dimension and makes the network shift invariant.
The second unit, Convolutional MLP (C-MLP) simply uses a depthwise convolution to transform the input.
The third kind, Temporal-Shift MLP (TS-MLP), concatenates shifted parts of the input.
We applied these MLP-based methods to the connectionist temporal classification (CTC)~\cite{Graves2006connectionist} ASR model and evaluated them on two datasets, Librispeech ~\cite{Panayotov2015librispeech} and Tedlium2 ~\cite{Rousseau2014ted2}.
We found trade-offs between accuracy in terms of word error rate (WER) and the number of parameters.
However, when matching the number of parameters, the best proposed MLP architectures outperform self-attention-based models, decreasing the WER by as much as \SI{1}{\percent} (or \SI{20}{\percent} relative improvement).

Our contributions in this work are as follows.
(i) We propose three MLP-based architectures suitable for sequences of arbitrary lengths.
(ii) We evaluate our architecture for ASR on two different datasets.
We show that the proposed C-MLP architecture outperforms the celebrated self-attention, both in accuracy and model size.
To the best of our knowledge, this is the first time that MLP-based architecture is applied to ASR.

\section{End-to-End ASR}
\subsection{Overview}
Recently, end-to-end (E2E) ASR models, where a single neural network produces a transcription directly from the input acoustic features, have come to dominate benchmarks.
The E2E ASR model consists of an encoder that converts the audio input into a latent representation and a decoder that converts the latent representation into text output (token).
To build the encoder-decoder architecture,
attention based encoder-decoder models such as Listen, Attend and Spell (LAS)~\cite{Chan15las} and recurrent neural network transducer (RNN-T)~\cite{Graves12rnnt} have shown high performance, 
but they suffer from slow inference speed due to the way they auto-regressively output tokens.
As a solution to the slow inference speed, non-autoregressive models with the audio encoder and a decoder using CTC~\cite{Graves2006connectionist} are attracting attention.

\subsection{CTC-based non-autregressive ASR}
In recent years, many studies have been conducted on CTC-based models and reported results that demonstrate high inference speed for a performance comparable to that of attention-based encoder-decoder models.
For the architecture of the audio encoder, recurrent neural networks (RNNs) and Convolution neural networks (CNNs) have been employed~\cite{graves2014towards, Li19jasper, Kriman2020Quartznet}.
More recently, CTC-based models applying the Transformer-based architecture~\cite{karita2019comparative, Zhang2020transformer-transducer, gulati2020conformer, Higuchi2021ACS} have shown promising performance and become the de facto architecture.

Transformer-based architecture has two components: point-wise feed-forward network and multi-head self-attention.
Point-wise feed forward network consists of MLP across the channel dimension.
Multi-head self-attention weighs the input signal based on the relationship between the input tokens. 
Transformer-based architecture first applies point-wise feed-forward network and then multi-head self-attention.
The input $\mathbf{X}_{in}$ processed as follows,
\begin{equation}
    \mathbf{X} = \mathrm{SelfAttention}(\mathbf{X_{in}})
\label{eq:FFN}
\end{equation}
\begin{equation}
    \mathbf{X_{out}} = \mathrm{FFN}(\mathbf{X})
\label{eq:MultiHeadAttn}
\end{equation}
where $\mathrm{SelfAttention}(\cdot)$ and $\mathrm{FFN}(\cdot)$ denote multi-head self-attention and point-wise feed-forward network.
Transformer-based models perform well because the relationship between the input tokens is captured effectively by self-attention.
However, it is of interest to investigate whether self-attention is truly necessary to capture this relationship.

\section{MLP-based architectures}
Next, we summarize the recently proposed MLP-based architectures.
The so-called MLP-based architectures are built using two modules, each based on simple MLPs: the channel mixing and token mixing modules.
The channel mixing module typically consists of linear projections on the channel dimension for each token, similar to the point-wise feed-forward network in the Transformer.
The token mixing module mixes the information in the token dimension as a replacement for multi-head self-attention in the Transformer, and how to design it is one of the keys to the MLP-based models.

\subsection{Conventional architectures}
MLP-Mixer~\cite{Tolstikhin21mlpmixer} consists of channel-mixing MLPs as channel-mixing module and token-mixing MLPs as token-mixing module.
It first applies token-mixing MLP.
The token-mixing MLP transposes the input $\mathbf{X_{in}} \in \mathbb{R}^{D \times N}$ and performs linear projection across the token dimension to mix the token information,
\begin{equation}
    \mathbf{X} =(\mathbf{W_2}\sigma(\mathbf{W_1}\mathbf{X_{in}^{\mathsf{T}}}))^{\mathsf{T}} \in \mathbb{R}^{D \times N},
\label{eq:mixer1}
\end{equation}
where $\mathbf{W_1} \in \mathbb{R}^{N^{\prime} \times N}$ and $\mathbf{W_2} \in \mathbb{R}^{N \times N^{\prime}}$ refer to weights of linear projections, and $\sigma$(·) is an element-wise activation function.
Albeit the components are the same as the channel-mixing module, the linear projections are done on the token dimension by transposing the input.
Next, the output of the token-mixing module, $\mathbf{X}$, is fed to the channel-mixing module.
The channel-mixing module is the same as the point-wise feed-forward network in the Transformer, which performs two linear projection across the channel dimension, 
\begin{equation}
    \mathbf{X_{out}} = \mathbf{W_4}\sigma(\mathbf{W_3}\mathbf{X}) \in \mathbb{R}^{D \times N}, 
\label{eq:mixer2}
\end{equation}
where $\mathbf{W_3} \in \mathbb{R}^{D^{\prime} \times D}$ and $\mathbf{W_4} \in \mathbb{R}^{D \times D^{\prime}}$ are the weights of the linear projections.
We remark that the construction of the MLP-Mixer type is similar to that of the Transformer, with the token-mixing replacing self-attention.
Equation~\ref{eq:mixer2} is equivalent to Equation~\ref{eq:FFN}.
This highlights the critical importance of token mixing.

In gMLP~\cite{Liu21gmlp}, the two layers of the channel-mixing module are split and the token-mixing module placed in-between.
As the token-mixing module, gMLP proposes a spatial gating unit (SGU), a module containing MLP across the token dimension and used to modulate the output of the unit.
Let $\mathbf{X_{in}} \in \mathbb{R}^{D \times N}$ denote the input data with $D$ channels and $N$ tokens.
First, linear projection on channel dimension $D$ to $D^\prime$ is applied to $\mathbf{X_{in}}$,
\begin{equation}
    \mathbf{X} =\sigma(\mathbf{W_1}\mathbf{X_{in}}) \in \mathbb{R}^{D^{\prime} \times N},
\label{eq:channel}
\end{equation}
where $\mathbf{W_1} \in \mathbb{R}^{D^{\prime} \times D}$ refers to weights of linear projection.
Next, the SGU follows.
The SGU first splits the input $\mathbf{X}$ into $\mathbf{X}_\mathrm{r} \in \mathbb{R}^{\frac{D^{\prime}}{2} \times N}$ and $\mathbf{X}_\mathrm{g} \in \mathbb{R}^{\frac{D^{\prime}}{2} \times N}$.
In gMLP, $\mathbf{X}_\mathrm{r}$ is set as the first half of $\mathbf{X}$ and $\mathbf{X}_\mathrm{g}$ for the second half of $\mathbf{X}$.
Then, it performs a gating operation.
The output of SGU $\mathbf{X}^\prime$ is,
\begin{equation}
    \mathbf{X}^\prime = \mathbf{X}_\mathrm{r} \odot \mathbf{H}_\mathrm{SGU} \in \mathbb{R}^{\frac{D^{\prime}}{2} \times N},
\label{eq:sgu}
\end{equation}
\begin{equation}
    \mathbf{H}_\mathrm{SGU} = \sigma((\mathbf{W_2}\mathbf{X^{\mathsf{T}}_\mathrm{g}})^\mathsf{T}) \in \mathbb{R}^{\frac{D^{\prime}}{2} \times N},
\label{eq:Hsgu}
\end{equation}
where $\odot$ denotes element-wise multiplication, and $\mathbf{W_2} \in \mathbb{R}^{N \times N}$ refers to the weights of the linear projection.
Finally, linear projection across the channel dimension $\frac{D^{\prime}}{2}$ to $D$ is applied as
\begin{equation}
    \mathbf{X_{out}} = \mathbf{W_3}\mathbf{X}^\prime \in \mathbb{R}^{D \times N},
\end{equation}
where $\mathbf{W_3} \in \mathbb{R}^{D \times \frac{D^{\prime}}{2}}$ is the weight matrix of the linear projection.
The effectiveness of the SGU has been experimentally shown with image and language experiments compared to non-gating linear projection and other variants.

S$^2$-MLP~\cite{Yu21s2mlp} and S$^2$-MLPv2~\cite{Yu21s2mlpv2} have a structure similar to gMLP and perform token-mixing by shift operation instead of the SGU.

GFNet~\cite{rao21gfnet} and FNet~\cite{Lee21fnet} use the discrete Fourier transform (DFT) in the token-mixing module.
GFNet efficiently applies learnable filters in the frequency domain in its token-mixing module.
It performs a 2D Fourier transform of the input image, multiplies by the filters, and transforms back to the image domain.
Then, MLP is applied across the channels in channle-mixing module.
FNet has been proposed in the area of NLP.
It applies a 2D Fourier transform to the input data and retains the real part in token-mixing module.
Then, it applies the MLP across channels in channel-mixing module.

\subsection{Challenges for ASR.} 

The sizes of the weight matrices in Equation~\ref{eq:mixer1} and~\ref{eq:Hsgu} are fixed and have to match the number of tokens in the input.
Similarly, the learnable filter in GFNet is of fixed size equal to the length of the input sequence.
These approaches are thus not suitable for variable length input.
Nevertheless, given a fixed dataset, they could be applied by zero-padding all sequences to the size of the longest one.
However, there are a number of shortcomings.
First and foremost, longer sequences that might be encountered at test time cannot be accommodated.
Then, MLPs will introduce a large number of parameters.
Moreover, they are not shift invariant, an important property for sequences.
While GFNet is shift invariant and only requires a reasonable number of parameters, it still cannot deal with sequences longer than the maximum size set.
FNet has neither of the above limitations, however, we find its performance somewhat lacking in the experiments of Section~\ref{sec:experiment}.

\begin{figure}
    \begin{center}
    \includegraphics{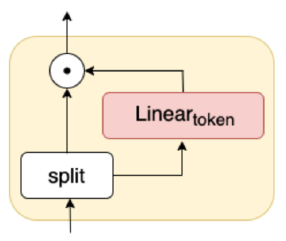}
    \end{center}
    \caption{The structure of the Spatial Gating Unit (SGU) in gMLP.}
    \label{fig:sgu}
\end{figure}

\begin{figure*}[t]
\begin{center}
\includegraphics[width=\linewidth]{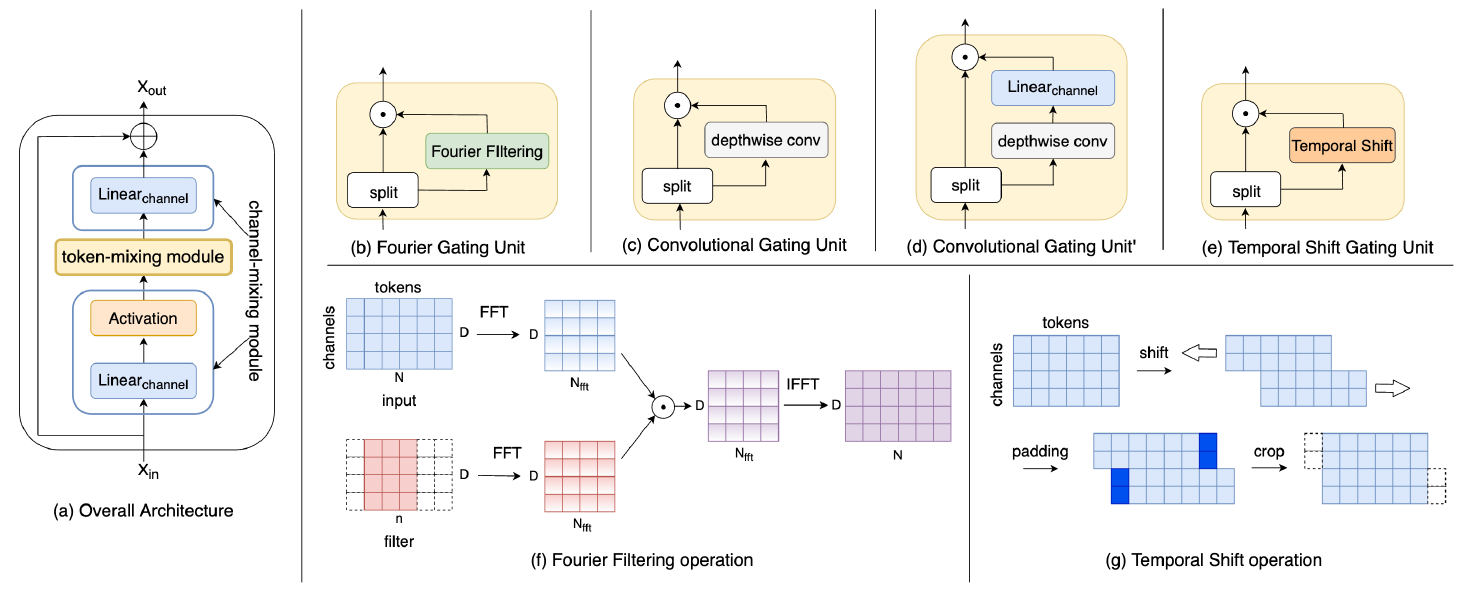}
\end{center}
\caption{Overview of (a) overall architecture, (b) Convolutional Gating Unit, (c) Convolutional Gating Unit$^{\prime}$, (d) Temporal-Shift Gating Unit, (e) Fourier Gating Unit (f) Temporal-Shift operation, and (g) Fourier Filtering.}
\label{fig:models}
\end{figure*}


\section{MLP-based architecture for Variable Length Input}
\label{sec:proposed_gating}

We describe here our three proposed MLP-based architectures for ASR.
All of our architectures are based on gMLP~\cite{Liu21gmlp}
They realize to take variable length input and shift invariant by replacing linear projection across the token dimension in SGU to other modules.
We first describe the overall architecture in \ref{sec:4-1}.
Then, we propose three token-mixing modules for ASR in \ref{sec:4-2}.
We show the name of our proposed architectures and token-mixing modules in Table~\ref{tbl:names}.

\subsection{Overall Architecture}
\label{sec:4-1}
Figure~\ref{fig:models}~(a) shows the overall architectures of our architectures where the two layers of the channel-mixing module are split and the token-mixing module placed in-between.
Let $\mathbf{X_{in}} \in \mathbb{R}^{D \times N}$ denote the input data with $D$ channels and $N$ tokens.
First, linear projection on channel dimension $D$ to $D^\prime$ is applied to $\mathbf{X_{in}}$,
\begin{equation}
    \mathbf{X} =\sigma(\mathbf{W_1}\mathbf{X_{in}}) \in \mathbb{R}^{D^{\prime} \times N},
\label{eq:channel}
\end{equation}
where $\mathbf{W_1} \in \mathbb{R}^{D^{\prime} \times D}$ refers to weights of linear projection.
Next, the token-mixing module follows,
\begin{equation}
    \mathbf{X}^\prime = \rm{TokenMixingModule}(\mathbf{X}) \in \mathbb{R}^{D^{\prime\prime} \times N}.
\label{eq:gmlp1}
\end{equation}
Finally, linear projection across the channel dimension $D^{\prime\prime}$ to $D$ is applied as
\begin{equation}
    \mathbf{X_{out}} = \mathbf{W_3}\mathbf{X}^\prime \in \mathbb{R}^{D \times N},
\end{equation}
where $\mathbf{W_3} \in \mathbb{R}^{D \times D^{\prime\prime}}$ is the weight matrix of the linear projection.

How to design the token-mixing module in Equation~\ref{eq:gmlp1} to realize variable-length input and shift invariant is the key to adapting MLP-based architecture to ASR.

\subsection{Token-mixing Module for Variable Length Input}
\label{sec:4-2}

\subsubsection{Fourier Gating Unit (FGU)}
The challenge in applying the conventional MLP-based architecture to ASR is that it cannot take variable length inputs.
The simplest solution is to align the sequence length by zero padding, but this is not shift invariant.
Therefore, we propose to perform zero padding in the time domain and then token-mixing in the frequency domain.
We call this token-mixing module Fourier Gating Unit (FGU) and refer to the overall architecture with FGU as Fourier-MLP (F-MLP).
The structure of FGU and its operation are illustrated in Figure~\ref{fig:models}~(b) and (f).
The fixed size filters are defined in the time domain, but applied in the frequency domain.
Thus, the filter is easily applied to sequences of any length by applying an FFT padded to the length of the input sequence to the filters, and working in the frequency domain.
The output of the FGU is computed as follows,
\begin{align}
    \mathbf{Z} = \mathcal{F}^{-1} \left[ \mathcal{F}[\mathbf{X}] \odot \mathcal{F}[\mathbf{K}] \right] \in \mathbb{R}^{D \times N},
\end{align}
where $\mathbf{X} \in \mathbb{R}^{D \times N}$ is the input signal and $\mathbf{K} \in \mathbb{R}^{D\times n}$ contains the filters in its rows.
The FFT and its inverse are denoted $\mathcal{F}[\cdot]$ and $\mathcal{F}^{-1}[\cdot]$, respectively.
We assume that the operation zero-pads $\mathbf{K}$ to size $D\times N$ to match $\mathbf{X}$.
This operation is equivalent to a circular depthwise convolution and shift invariant.

We also considered defining the MLP in the frequency domain instead of the time domain filter just described.
But preliminary experiments showed only modest performance for a large number of parameters.

GFNet~\cite{rao21gfnet} also applies element-wise multiplication to learnable filters and input features in the frequency domain.
To apply GFNet to variable sequence length input, we have to find the maximum sequence length of the dataset and define filters to match that length.
On the other hand, FGU defines filters in the time domain, zero-pads it to match the size of each input data, and performs FFT to the filter and input data.
Therefore, F-MLP has fewer parameters than GFNet.

\subsubsection{Convolutional Gating Unit (CGU)}
\label{sec:cmlp}
\textbf{Convolutional Gating Unit (CGU).}
FGU uses FFT-based circular depthwise convolution to mix the temporal information.
For the second approach, we use a convolutional layer along the time dimension.
We call this token-mixing module Convolutional Gating Unit (CGU) and refer to the overall architecture with CGU as Convolutional-MLP (C-MLP).
The structure of a CGU is shown in Figure~\ref{fig:models}~(c).
CGU replaces linear projection across the token dimension in SGU (see Equation~\ref{eq:Hsgu}) with a depthwise convolution.
Its output $\mathbf{H}_\mathrm{CGU}$ is,
\begin{equation}
    \mathbf{H}_\mathrm{CGU} = \mathbf{K} \star \mathbf{X}_\mathrm{g} \in \mathbb{R}^{\frac{D}{2} \times N}, 
\label{eq:Hcgu} \\
\end{equation}  
where $\mathbf{X}_\mathrm{g}  \in \mathbb{R}^{\frac{D}{2} \times N}$, $\mathbf{K}  \in \mathbb{R}^{\frac{D}{2} \times k}$ is the $\frac{D}{2}$-dimensional kernel with kernel size $k$.
The depth wise convolution operation is denoted by $\star$ and defined as, 
\begin{equation}
    (\mathbf{K} \star \mathbf{X})_{:,i} = \sum_{j=1}^{k} \mathbf{K}_{:,j} \odot \mathbf{X}_{:,k+i-j}.
\label{eq:depthwise}
\end{equation}  

\textbf{Convolutional Gating Unit$^{\prime}$ (CGU$^{\prime}$).}
We also propose a variation where a linear projection across the channel dimension is applied to the filter $\mathbf{H}_\mathrm{CGU}$.
In this case, information from both token and channel dimensions is mixed.
The new filter $\mathbf{H}^{\prime}_\mathrm{CGU}$ is formulated as follows,
\begin{equation}
    \mathbf{H}^{\prime}_\mathrm{CGU} =  \mathbf{W}\mathbf{H}_\mathrm{CGU} \in \mathbb{R}^{\frac{D}{2} \times N}.
\end{equation}
where $\mathbf{W} \in \mathbb{R}^{\frac{D}{2} \times \frac{D}{2}}$ is the weight matrix of the linear projection.
This structure is illustrated in Figure~\ref{fig:models}~(c).
We refer to this token-mixing module as Convolutional Gating Unit$^{\prime}$ (CGU$^{\prime}$) and overall architecture with (CGU$^{\prime}$) as  Convolutional-MLP$^{\prime}$ (C-MLP$^{\prime}$).

\subsubsection{Temporal-shift Gating Unit (TSGU)}
In CGU, we used learnable kernels for convolution across the token dimension.
Next, we consider the case of performing a convolution using kernels with fixed parameters.
If the input is a $D\times N$ matrix, then the $i$th row of the fixed kernel $\mathbf{K}_{\text{shift}}$ is
\begin{equation}
\mathbf{k}_{i} = \begin{cases}
[0, 0, 0, 0, 1] & \text{if $1 \leq i \leq \frac{D}{2}$,} \\
[1, 0, 0, 0, 0] & \text{if $\frac{D}{2} < i \leq D$.}
\end{cases}
\label{eq:tsgu}
\end{equation}
This corresponds to the shift operation proposed in S$^2$-MLP~\cite{Yu21s2mlp}.
Half of the channels are shifted by two forward, and the other half by two backward along the time dimension.
The time shift operation of this kernel on the input signal is illustrated in Figure~\ref{fig:models}~(f).
The size of the shift is a parameter but we fix it here to the value taken in the experiments.
We propose the Temporal Shift Gating Unit (TSGU) as a token-mixing module with a shift operation in the token direction.
We refer to the overall architecture with TSGU as Temporal Shift-MLP (TS-MLP).
The inputs is split as in SGU, and the gating values derived as
\begin{equation}
    \mathbf{H}_\mathrm{TSGU} = \mathbf{K}_{\text{shift}} \star \mathbf{X}_g,
\end{equation}  
which replaces $\mathbf{H}_\mathrm{SGU}$.
This gating mechanism is parameter-free and mixes time information by applying progressive time shifts to the input signal.

\begin{table}[t]
\caption{Overview of the proposed method.}
\label{tbl:names}
\begin{center}
\footnotesize
\begin{tabular}{@{}lccc@{}}
\toprule
\textbf{model} & \textbf{token-mixing} & \textbf{method} \\
\midrule
F-MLP             &FGU             &circular depthwise convolution \\ 
C-MLP             &CGU             &depthwise convolution \\ 
C-MLP$^{\prime}$  &CGU$^{\prime}$  &   + linear projection \\ 
TS-MLP            &TS-GU           &temporal shift operation \\
\bottomrule 
\end{tabular}
\end{center}
\end{table}

\section{Experiments}
\label{sec:experiment}

\subsection{Experimental Setup}

We applied our architectures to non-autoregressive CTC-based ASR.
The experiments were conducted using the ESPNet tool kit~\cite{Watanabe2018espnet}.
We use Transformer-encoder as self-attention-based baseline, which is a strong baseline used in many prior works~\cite{Bai2020ListenAA, Higuchi2021ACS, lee2021intermediate}.
In addition, we evaluate FNet~\cite{Lee21fnet} and GFNet~\cite{rao21gfnet} as MLP-based baseline.
In order to apply GFNet to variable length data, it is necessary to set the filter size to the maximum sequence length.
Therefore, GFNet cannot process sequences longer than this.
In order to justify the performance of the architecture itself, all parameters other than the architecture were kept the same in all experiments.

\begin{table*}[t]
\caption{Experimental results on Librispeech and Tedlium2.}
\label{tbl:wer}
\begin{center}
\begin{tabular}{@{}lrcccccc@{}}
\toprule
\textbf{} & \multicolumn{1}{l}{\textbf{}} &\multicolumn{4}{c}{\textbf{WER (Librispeech)}} & \multicolumn{2}{c}{\textbf{WER (Tedlium2)}} \\
\textbf{method} & \textbf{params} & \textbf{dev-clean} & \textbf{dev-other} & \textbf{test-clean} & \textbf{test-other} & \textbf{dev} & \textbf{test} \\
\midrule
\multicolumn{6}{@{}l@{}}{\emph{Baseline}} \\
\quad Transformer-based model    &16.2M &6.7\% &16.3\% &6.7\% &16.2\% &14.4\% &13.9\% \\
\quad FNet~\cite{Lee21fnet}     &11.5M &15.7\% &31.3\% &15.8\% &31.9\% &31.4\% &30.4\% \\
\quad GFNet~\cite{rao21gfnet}   &16.2M &5.1\% &13.6\% &5.4\% &13.8\% &13.2\% &12.6\% \\
\midrule
\multicolumn{6}{@{}l@{}}{\emph{Ours}} \\
\quad F-MLP                      &9.2M &9.0\% &21.8\% &9.4\% &22.2\% &19.3\% &18.3\% \\
\quad C-MLP                      &9.3M &6.3\% &16.7\% &6.3\% &16.8\% &14.6\% &13.6\% \\
\quad C-MLP$^{\prime}$           &14.0M &5.7\% &15.4\% &5.8\% &15.7\% &13.6\% &12.8\% \\
\quad TS-MLP                     &\textbf{9.1}M  &8.9\% &22.1\% &8.7\% &22.8\% &19.2\% &18.4\% \\
\quad F-MLP+tiny attn            &12.2M &6.4\% &16.1\% &6.5\% &16.4\% &15.3\% &14.9\% \\
\quad C-MLP+tiny attn            &12.2M &5.2\% &13.8\% &5.5\%  &13.8\% &12.8\% &12.3\% \\
\quad C-MLP$^{\prime}$+tiny attn &16.9M &\textbf{4.8}\% &\textbf{12.8}\% &\textbf{5.2}\% &\textbf{13.1}\% &\textbf{12.3}\% &\textbf{11.7}\% \\ 
\quad TS-MLP+tiny attn           &12.1M &6.3\% &16.3\% &6.4\% &16.3\% &14.5\% &14.2\% \\
\bottomrule
\end{tabular}
\end{center}
\end{table*}

\textbf{Encoder layer structure.}
For our proposed architecture, input and output dimensions of the first channel-mixing module are set to 256 and 1024, input and output dimensions of the token-mixing module are set to 512 and 256, input dimension and output dimension of the second channel-mixing module are set to 1024, 256.
For FNet and GFNet, input and output dimensions of the token-mixing module are set to 256, input dimension, hidden dimension, and output dimension of the channel-mixing module are set to 256, 1024, 256.
For every architecture,  we use Gaussian Error Linear Units (GELU)~\cite{DAN16gelu} as the activation function.
Convolution kernel size of C-MLP and C-MLP$^{\prime}$ and filter size of F-MLP are 15 and shift size of TS-MLP is 2.
For Transformer encoder, we follow the architecture proposed in~\cite{karita2019improving}.
The number of heads and input dimensions of the self-attention module are 4 and 256.
The intermediate dimensions of the feed-forward network are 1024.
For all models, we use two CNN-based subsampling layers before encoder layers.
We set the subsampling rate to 0.25.
We experiment with all the models in an 18-layer setup.
We also experiment with the case where all model sizes are scaled to the same.

\textbf{Data.}
We measure performance on two datasets: Librispeech and Tedlium2.
Librispeech contains 960-hour utterance from read English audiobooks.
Specifically, the validation and test sets of Librispeech are divided into “clean” and “other” based on the quality of the recorded utterances. 
Tedlium2 contains utterances from English Ted Talks, and we used the 207-hour training data. 

We use 80-dimensional log-mel-spectrogram features and 3-dimensional pitch features as inputs.
For feature extraction, we use kaldi~\cite{Povey2011kaldi},  and shift size and length of the window are set to \SI{10}{\milli\second} and \SI{25}{\milli\second}, respectively.
We apply SpecAugment~\cite{Park19specaugment} and speed perturbations~\cite{Ko15audio} for data augmentation.
For Librispeech, we tokenize text into 300 subwords, and for Tedlium2, we tokenize text into 500 subwords.
We created subwords with SentencePiece ~\cite{kudo18sentencepiece} 

\textbf{Training and inference setup.}
All models are trained for 50 epochs.
We set the batch size to 64.
The optimizer is Adam with $\beta_1=0.9$, $\beta_2=0.98$, $\epsilon =10^{-9}$.
The scheduling method for the learning rate is the same as~\cite{vaswani2017attention}, 
$
\mathrm{learning\: rate} = d^{-0.5} \cdot \min(\mathrm{step\_num}, \mathrm{step\_num} \cdot \mathrm{warmup\_steps}^{-1.5}),
$
where we set $\mathrm{warmup\_steps}$ and $d$ to 25000 and 1280, respectively.
Dropout rate and label smoothing rate are set to 0.1.
For inference, we use the model parameters obtained by averaging the 10 models with the best validation scores.
The outputs are decoded by greedy decoding for CTC, without using any external language model.

\begin{figure}[t]
\begin{center}
\includegraphics{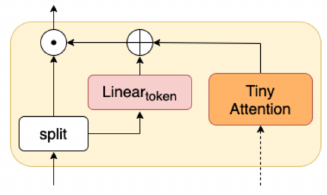}
\end{center}
\caption{SGU with a tiny self-attention module.}
\label{fig:amlp}
\end{figure}

\textbf{Tiny self-attention module.}
gMLP~\cite{Liu21gmlp} demonstrate that adding a small self-attention module to the SGU can improve performance at the cost of a modest increase in resource usage.
The structure of an SGU with a tiny self-attention module is shown in Figure~\ref{fig:amlp}.
The input of the tiny self-attention module is the input of the encoder.
The output of the tiny self-attention module is added to the end of the right path of SGU.
The tiny self-attention module has the same structure as the self-attention module in the Transformer encoder, but its hidden dimension of linear projection $d$ and the number of attention heads $n_{\mathrm{head}}$ are small.
We also experimented with the proposed token-mixing module combined with tiny attention.
We set the tiny self-attention to $n_{\mathrm{head}}=1, d=128$, while we set self-attention module in the Transformer-based model to $n_{\mathrm{head}}=4, d=256$.

\begin{table*}[t]
\caption{WER on the same large model size. In order to stabilize the learning of deep models, all models were trained using InterCTC~\cite{lee2021intermediate}, a regularization technique used in CTC-based ASR.}
\label{tbl:scale}
\begin{center}
\begin{tabular}{@{}lrccccc@{}}
\toprule
\textbf{method} & \textbf{layers} & \textbf{params} & \textbf{dev-clean} & \textbf{dev-other} & \textbf{test-clean} & \textbf{test-other} \\
\midrule
Transformer-based model L    &36 &30.4M &4.2\% &11.4\% &4.5\% &11.3\% \\
F-MLP L                      &72 &31.0M &5.4\% &14.6\% &5.8\% &14.5\% \\
C-MLP L                      &72 &31.1M &3.4\% &9.9\% & 3.5\% &9.9\% \\
C-MLP$^{\prime}$ L           &42 &30.0M &3.6\% &10.9\% &3.8\% &10.7\% \\
TS-MLP L                     &72 &30.5M &4.6\% &13.2\% &4.7\% &13.4\% \\
F-MLP+tiny attn L            &51 &31.0M &\textbf{3.3}\% &9.4\% &3.6\% &9.3\% \\
C-MLP+tiny attn L            &51 &31.0M &\textbf{3.3}\%  &\textbf{8.6}\% &\textbf{3.4}\% &\textbf{8.6}\% \\
C-MLP$^{\prime}$+tiny attn L &33 &29.4M &\textbf{3.3}\% &9.1\% &3.6\% &9.1\% \\
TS-MLP+tiny attn L           &51 &30.6M &3.7\% &9.6\% &3.8\% &9.7\%\\
\bottomrule
\end{tabular}
\end{center}
\end{table*}

\subsection{Results}
\label{sec:result}
\textbf{Main results.} 
Table~\ref{tbl:wer} provides a comparison of the parameter sizes and word error rates (WER) on Librispeech and Tedlium2.
In Table~\ref{tbl:wer}, we see that C-MLP achieves competitive performance with Transformer-based model with only 57.4\% of its parameters.
C-MLP$^{\prime}$ achieves the best WER in Table~\ref{tbl:wer} and improves WER by 1.0 / 0.9\%, 0.9 / 0.5\% on Librispeech dev-clean/dev-other, test-clean/test-other set, and 0.8 / 1.1\% on Tedlium2 dev/test set.
It increases the model size a little but is still only 86.4\% the size of Transformer-based model.
TS-MLP, which can be said to be a special case of C-MLP, has the smallest number of parameters.
TS-MLP has only 56.2\% of parameters of Transformer-based CTC while degrading WER by 2.2 / 5.8\%, 2.0 / 6.6\% on Librispeech dev-clean/dev-other, test-clean/test-other set and 4.8 / 4.5\% on Tedlium2 dev/test set.
F-MLP is 56.8\% the size of GFNet, which also uses Fourier transform, and Transformer-based model.
Compared to the Transformer-based model, F-MLP degrades WER by 2.3 / 5.5\%, 2.7 / 6.0\% on Librispeech dev-clean/dev-other, test-clean/test-other sets and 4.9 / 4.4\% on Tedlium2 dev/test sets.

\textbf{Results with a tiny attention module.} 
Furthermore, in Table~\ref{tbl:wer}, we can see that hybrid models of the self-attention and MLP using a tiny self-attention module can greatly improve the performance.
This suggests that token-mixing by self-attention and the proposed methods complement each other well.
Here we show that the combination is effective, even when the tiny self-attention module has reduced parameter size compared to the fully Transformer-based module.
When combined with the tiny self-attention module, TS-MLP and F-MLP achieve competitive performance with the Transformer-based model while they degrade WER on their own.
The number of parameters is a little larger, but still about \SI{75}{\percent} of that of the Transformer.
C-MLP and C-MLP$^{\prime}$ also improve WER when combined with a tiny self-attention module.
C-MLP$^{\prime}$ with a tiny attention achieves best performance. Compared with the Transformer-based model, it improves WER by 1.9 / 3.5\%, 1.5 / 3.3\% on Librispeech dev-clean/dev-other, test-clean/test-other set, and 2.1 / 2.2\% on Tedlium2 dev/test set with almost the same number of parameters.

\textbf{Results on the same large model size.}
Table~\ref{tbl:scale} shows the performance when the model is scaled so that the number of parameters is about 30M.
We can see that the performance is improved by increasing the number of layers, comparing the results in table~\ref{tbl:wer} and table~\ref{tbl:scale}.
Our models have a smaller number of parameters than Transformer-based model, so more layers can be added.
Among our architectures, C-MLP achieves the best performance, with a 0.8 / 1.5\%, 1.0 / 1.4\% improvement in WER on Librispeech dev-clean/dev-other, test-clean/test-other set compared to the Transformer-based model.
We can also sett that our architectures improve their performance by combining with a tiny self-attention module.
Compared to the results in Table~\ref{tbl:wer}, C-MLP, C-MLP$^{\prime}$, and TS-MLP show a larger performance improvement than the other models.
This may be because the number of layers of them is larger than other models due to their smaller size.
We conjecture that the larger number of layers allows better mixing between a wider range of locations, leading to improved representations.

\begin{table}[t]
\caption{Computational cost and the number of parameters of linear projection across the channel dimension and token dimension, self-attention, tiny-self-attention, and proposed token-mixing module. $N$ is the sequence length and $D$ is the size of channel dimension of a feature. $l$ is the size of filters in Fourier Filter Unit. $k$ is the convolution kernel size.}
\label{tbl:models}
\begin{center}
\footnotesize
\begin{tabular}{@{}lccc@{}}
\toprule
\textbf{Method} & \textbf{computational complexity} & \textbf{parameters} \\
\midrule
linear (channel)        &$ND^2$             &$D^2$ \\ 
linear (token)          &$N^2D$             &$N^2$ \\ 
self-attention          &$4ND^2+2N^2D$      &$4D^2$ \\ 
\midrule
FGU                     &$ND \log_2 N+ND$   &$lD$ \\
CGU                     &$kND^2+ND$         &$kD$ \\
TSGU                    &$N+ND$             &-  \\
tiny self-attention     &$2ND^2+2N^2D$      &$2D^2$ \\ 
\bottomrule 
\end{tabular}
\end{center}
\end{table}

\textbf{Model size and Computational Cost Analysis.}
We show computational characteristics of each architecture in Table~\ref{tbl:models}.
$N$ denotes the input sequence length and $D$ the number of channels.
$l$ is the size of filters in Fourier Filter Unit and $k$ is the convolution kernel size.
Since $lD<N$ in general, FGU in F-MLP has smaller computational complexity and fewer parameters than MLP in the token direction and self-attention.
In the case of GFNet~\cite{rao21gfnet}, we have to set l to the maximum length of the dataset, but for FGU, it can be a smaller.
CGU in C-MLP has the convolution layer, so the computational complexity is larger but the number of parameters is smaller than self-attention.
TS-MLP, which is a special case of C-MLP, only performs a simple shift operation and element-wise multiplication. 
The TSGU has no parameters at all, and achieves the smallest computational complexity, and fewest parameters, in Table~\ref{tbl:models}.
The tiny self-attention module has only a single head and smaller hidden dimensions, so the computational cost and number of parameters are smaller than in the Transformer-based model.

\section{Conclusion}



We proposed three new network architectures based on MLP-mixing for sequences of variable size.
Each uses a different mechanism to mix information across the tokens and obtain shift invariance.
F-MLP simply relies on gating with circular convolution, C-MLP on convolutional gating, and TS-MLP on time-shift gating.
Extensive experiments revealed that these MLP architectures are sufficient to outperform Transformer-based models for non-autoregressive ASR.
Among the different models, C-MLP was the best.
Although the proposed architectures alone can provide sufficient performance, by adding a small self-attention module, we can improve the performance while keeping the number of parameters low.
We thus conclude that all three proposed MLP-like architectures are not only suitable for ASR, but also highly practical due to their simplicity and good performance.
In the future, we will explore their application to other tasks such as natural language processing or acoustic event detection.




\bibliographystyle{IEEEbib}
\bibliography{strings,refs}

\newpage
\appendix
\onecolumn
\section{Appendix: Hyperparameters for our experiments}
We summarize the hyperparameters described in section~\ref{sec:experiment} in Table~\ref{tbl:params} and Table~\ref{tbl:dims}.

\begin{table}[h]
\caption{Hyperparameters}
\label{tbl:params}
\begin{center}
\footnotesize
\begin{tabular}{@{}llc@{}}
\toprule
&\textbf{hyperparameter} &\textbf{value}\\
\midrule
&Epoch      &50 \\
&batch size &64 \\
&dropout rate           &0.1 \\
&label smoothing rate   &0.1 \\
&subsampling rate       &0.25\\
&Optimizer              &Adam($\beta_1 = 0.9, \beta_1 = 0.98, \epsilon = 10^{-9}$) \\
&learning rate          &$d^{-0.5} \cdot \min(\mathrm{step\_num}, \mathrm{step\_num} \cdot \mathrm{warmup\_steps}^{-1.5})$ \\
&warmup steps       &25000\\
&$d$       &1280\\
&window length (Feature Extraction) &25ms \\
&window shift (Feature Extraction) &10ms \\
\bottomrule
\end{tabular}
\end{center}
\end{table}

\begin{table}[h]
\caption{input and output channel dimensions of the architectures}
\label{tbl:dims}
\begin{center}
\footnotesize
\begin{tabular}{@{}llc@{}}
\toprule
\textbf{} &\textbf{module} &\textbf{parameters}\\
\midrule
\multirow{4}{*}{Transformer encoder}
&self-attention &$\mathrm{num\_head}=4, \mathrm{hidden dim=256}$ \\
&Linear$^{(1)}$ in FFN &$\text{input}=256, \text{output}=1024$ \\
&Linear$^{(2)}$ in FFN &$\text{input}=1024, \text{output}=256$ \\
&Activation             &GELU \\
\midrule
\multirow{6}{10em}{GFNet, FNet}
&token-mixing module      &$input=256, output=256$ \\
&Linear$^{(1)}_{\mathrm{channel}}$ &$\text{input}=256, \text{output}=1024$ \\
&Linear$^{(2)}_{\mathrm{channel}}$ &$\text{input}=1024, \text{output}=256$ \\
&Activation             &GELU \\
&filter size (GFNet) &512 \\
&tiny self-attention &$\mathrm{num\_head}=1, \mathrm{hidden dim=128}$ \\
\midrule
\multirow{8}{12em}{Ours}
&Linear$^{(1)}_{\mathrm{channel}}$ &$\text{input}=256, \text{output}=1024$ \\
&token-mixing module      &$\text{input}=1024, \text{output}=512$ \\
&Linear$^{(2)}_{\mathrm{channel}}$ &$\text{input}=512, \text{output}=256$ \\
&Activation             &GELU \\
&filter size (F-MLP) &15 \\
&convolution kernel size (C-MLP, C-MLP$^{\prime}$) &15 \\
&shift size (TS-MLP) &2 \\
&tiny self-attention &$\mathrm{num\_head}=1, \mathrm{hidden dim=128}$ \\
\bottomrule
\end{tabular}
\end{center}
\end{table}

\end{document}